%% file: main.tex
\documentclass{WileyMSP-template}
\usepackage{graphicx}%
\usepackage{multirow}%
\usepackage{amsmath,amssymb,amsfonts}%
\usepackage{amsthm}%
\usepackage{mathrsfs}%
\usepackage[title]{appendix}%
\usepackage{xcolor}%
\usepackage{textcomp}%
\usepackage{manyfoot}%
\usepackage{booktabs}%
\usepackage{algorithm}%
\usepackage{algorithmicx}%
\usepackage{algpseudocode}%
\usepackage{listings}%
\usepackage{array}
\usepackage{bm}
\usepackage{hyperref}
\hypersetup{hidelinks,
	colorlinks=true,
	allcolors=black,
	pdfstartview=Fit,
	breaklinks=true}
\usepackage{marvosym}
\input{math_command}

\begin{document}

\pagestyle{fancy}
\rhead{\includegraphics[width=2.5cm]{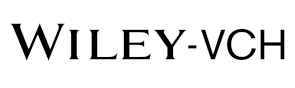}}

\title{End-to-End Crystal Structure Prediction from Powder X-Ray Diffraction}

\maketitle


\author{Qingsi Lai}
\author{Fanjie Xu}
\author{Lin Yao*}
\author{Zhifeng Gao}
\author{Siyuan Liu}
\author{Hongshuai Wang}
\author{Shuqi Lu}
\author{Di He}
\author{Liwei Wang}
\author{Linfeng Zhang}
\author{Cheng Wang*}
\author{Guolin Ke*}


\dedication{}

\begin{affiliations}
Q. Lai, F. Xu, L. Yao, Z. Gao, S. Liu, H. Wang, S. Lu, L. Zhang, G. Ke\\
DP Technology, Beijing, 100080, China\\
Email Address: yaol@dp.tech; kegl@dp.tech\\

Q. Lai, L. Wang\\
Center for Data Science, Peking University, Beijing 100871, China \\

D. He, L. Wang\\
School of Intelligence Science and Technology, Peking University, Beijing 100871, China\\

F. Xu, C. Wang\\
College of Chemistry and Chemical Engineering, Xiamen University, Xiamen, 361005, China\\
Email Address: wangchengxmu@xmu.edu.cn\\

L. Zhang, C. Wang\\
AI for Science Institute, Beijing, 100084, China

\end{affiliations}


\keywords{Deep Learning, Powder X-ray Diffraction, Crystal Structure Prediction, Equivariant
Deep Generative Model, Metal-organic Frameworks (MOFs)}

\begin{abstract}

Powder X-ray diffraction (PXRD) is a prevalent technique in materials characterization. While the analysis of PXRD often requires extensive human manual intervention, and most automated method only achieved at coarse-grained level. The more difficult and important task of fine-grained crystal structure prediction from PXRD remains unaddressed. This study introduces XtalNet, the first equivariant deep generative model for end-to-end crystal structure prediction from PXRD. Unlike previous crystal structure prediction methods that rely solely on composition, XtalNet leverages PXRD as an additional condition, eliminating ambiguity and enabling the generation of complex organic structures with up to 400 atoms in the unit cell.  XtalNet comprises two modules: a Contrastive PXRD-Crystal Pretraining (CPCP) module that aligns PXRD space with crystal structure space, and a Conditional Crystal Structure Generation (CCSG) module that generates candidate crystal structures conditioned on PXRD patterns.  Evaluation on two MOF datasets (hMOF-100 and hMOF-400) demonstrates XtalNet's effectiveness. XtalNet achieves a top-10 Match Rate of 90.2\% and 79\% for hMOF-100 and hMOF-400 in conditional crystal structure prediction task, respectively. XtalNet enables the direct prediction of crystal structures from experimental measurements, eliminating the need for manual intervention and external databases. This opens up new possibilities for automated crystal structure determination and the accelerated discovery of novel materials.

\end{abstract}


\section{Introduction}\label{sec1}

Powder X-ray Diffraction (PXRD)~\cite{holder2019tutorial} is widely used material characterization techniques for its cost-effectiveness and its ability to analyze crystallography~\cite{dinnebier2008powder}. PXRD produces a pattern that encodes information about the crystal symmetry, lattice parameters and atomic positions~\cite{de2012structure}. Comparing XRD patterns of candidate materials with the measured or simulated XRD patterns of known materials allows for deriving all the aforementioned information. However, the analysis of PXRD patterns often involves multiple sequential steps and extensive manual intervention, leading to several attempts to automate PXRD analysis~\cite{chen2021automating,szymanski2023adaptively,maffettone2021crystallography, ParkClassification, oviedo2019fast, SuzukiSymmetry}. Although these efforts have achieved success at coarse-grained levels, such as phase identification~\cite{szymanski2023adaptively} and space group prediction~\cite{SuzukiSymmetry}, the more challenging task of fine-grained crystal structure prediction from PXRD remains unaddressed. In this paper, we propose an end-to-end method named XtalNet to identify crystal structures from PXRD.

In PXRD analysis, determining the crystal structure of experimental materials is one of the most important and challenging steps. Existing PXRD-based crystal structure determination methods involve comparing experimental PXRD patterns of unknown materials against databases like the Inorganic Crystal Structure Database (ICSD)~\cite{icsd} to identify structural analogues as a starting point. Subsequently, Rietveld refinement\cite{Rietveld1969APR} is employed to incrementally refine the rough analogous structure and achieve greater precision. However, the incompleteness of the database and the complexity of the procedures complicate the structure determination process. Additionally, the Rietveld refinement step often requires extensive manual intervention from experienced scientists. Therefore, a method that can directly predict crystal structures from PXRD data without relying on external databases has the potential to significantly reduce the manual labor required by chemists and represent a major stride towards automating PXRD analysis.

Crystal structure plays a critical role in various scientific realms such as physics, chemistry, and material science, as many physical and chemical properties of materials are determined by the crystal structure~\cite{butler2018machine,xie2018crystal,oganov2011evolutionary}. Crystal structure prediction (CSP), which aims to obtain the three-dimensional structure of crystals based on their composition~\cite{desiraju2002cryptic}, has made significant progress with machine learning~\cite{wengert2021data,court20203,hu2021contact,dan2020generative,cdvae,diffcsp,zeni2023mattergen,cheng2022crystal}. Most of these methods are carefully designed to account for the unique challenges of periodicity and equivariance in CSP, and some of them~\cite{cdvae,diffcsp,zeni2023mattergen} utilize diffusion models~\cite{DDPM,song2020score} to achieve remarkable structure generation results. However, these methods have been primarily tested on inorganic crystal structure datasets such as Perov-5~\cite{CastelliNew,CastelliComputational}, MP-20~\cite{JainCommentary}, and ICSD~\cite{zagorac2019recent}, which consist of a limited number of atoms in the unit cell (usually less than 50). The efficacy of these methods in predicting more complex and larger organic structures or under stringent conditions has yet to be validated. Furthermore, the CSP problem is typically formulated to predict the minimum energy structure. However, as the number of atoms in a unit cell increases, one composition may correspond to many crystal structures as the energy landscape becoming complex. Therefore, a more practical scenario is predicting a material's crystal structure based on its experimental characterization. The existence of this characterization leads to a one-to-one correspondence with crystal structure. Thus, integrating experimental characterization into crystal structure prediction is a more practical application scenario. Previous crystal structure prediction methods provide us with a powerful tool for generating crystal structures. Based on this powerful tool, what we need to consider is how to use experimental characterization as a condition to guide the crystal structure prediction.

While PXRD is fundamentally different from crystal structure, a sole prediction module is not able to achieve finer structure prediction that satisfies the PXRD condition, as shown in Results~\ref{diff_arch}. A prediction module is required to predict the crystal structure and align the PXRD latent space with the crystal space simultaneously, which poses a significant challenge. Therefore, incorporating prior information into the prediction module to provide alignment between the PXRD space and crystal structure space is necessary. A common method is to use pretraining, which obtains a post-aligned PXRD encoder as the initialization of the prediction module. The necessity of the pretraining procedure for multi-modality tasks has been validated in many other fields, such as text-to-image generation~\cite{saharia2022photorealistic,ruiz2023dreambooth,gal2022image,zhang2023adding}, with prominent examples such as Stable Diffusion~\cite{stablediffusion}. These models employ text embeddings generated by pretrained CLIP~\cite{CLIP} text encoders to guide the diffusion process in generating images conditioned on textual descriptions. In the case of crystal structure prediction, PXRD serves as the key signal to guide the generation of crystal structures. Drawing inspiration from these successful multi-modality models in other domains, we have explored the potential of integrating a contrastive learning approach into our conditional crystal structure prediction methodology, resulting in the contrastive PXRD-crystal pre-training (CPCP) module. By pretraining the PXRD encoder using the Contrastive PXRD-Crystal Pre-training (CPCP) module, PXRD signals closer to the crystal structure space can be provided. The PXRD encoder pretraining by the CPCP module can guide the generation of crystal structures.

Therefore, we propose \textbf{XtalNet}, the \textit{first} equivariant deep generative model for end-to-end crystal structure prediction from PXRD. XtalNet aims to extend the capabilities of deep learning in predicting crystal structures based on PXRD patterns, encompassing more complex structures and specific conditions. Unlike previous CSP methods that solely rely on composition for obtaining crystal structures, XtalNet incorporates PXRD as a supplemental condition, ensuring a one-to-one mapping without ambiguity, which can be applied in a real experimental setting. To the best of our knowledge, XtalNet is the first deep learning method to directly predict the fine-grained crystal structure, which makes it different from previous PXRD analysis methods that only predict coarse-grained properties. XtalNet is an end-to-end deep learning based method that eliminates the need for human intervention or external database dependencies, distinguishing it from traditional PXRD analysis approaches involving complex multi-step procedures and human insights. Our model is capable of handling organic crystal systems, such as metal-organic frameworks (MOFs) that are important for gas separation~\cite{knebel2022metal}, even when the unit cell contains a large number of atoms (up to 400). To achieve this, we have compiled PXRD-crystal datasets named hMOF-100 and hMOF-400 with more than 100,000 data points.

Our approach is made possible through the contrastive PXRD-crystal pre-training (CPCP) module and the conditional crystal structure generation (CCSG) module. The CPCP module employs contrastive learning pre-training to align the PXRD space with the crystal structure space. Furthermore, by utilizing the CCSG modules, multiple candidate crystal structures can be generated conditioned on the PXRD pattern by equivariant diffusion model. These candidate structures are subsequently scored and ranked using the CPCP module, thus accomplishing the ranked structure prediction task with a top-10 match rate of 90.2\% in the hMOF-100 dataset and 79\% in the hMOF-400 dataset.

\section{Results}

\subsection{Overview of XtalNet}
\begin{figure}[thbp!]
    \centering
    \includegraphics[width=0.95\textwidth]{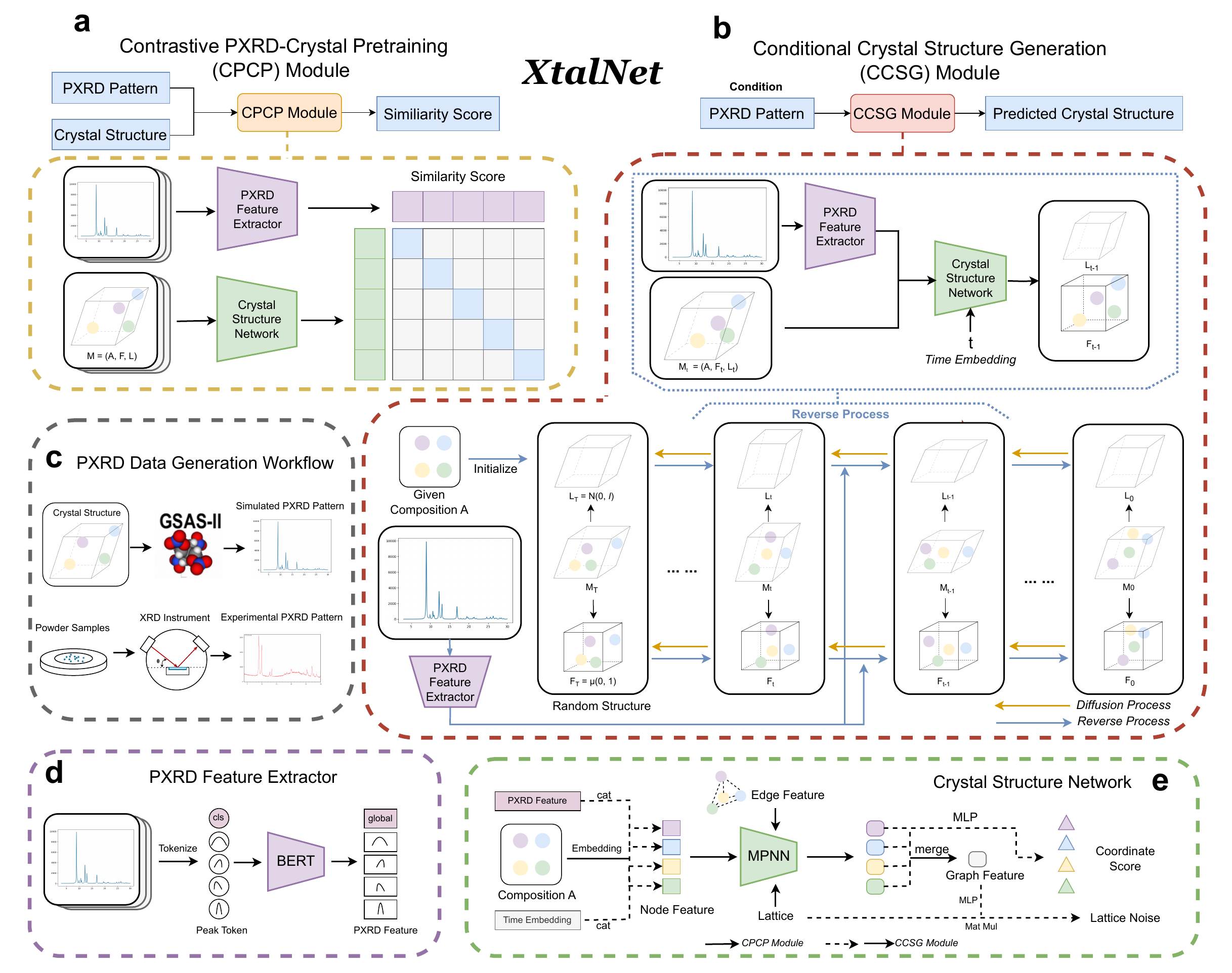}
    \caption{\textbf{Overview of XtalNet.} \textbf{a}, Framework of the Contrastive PXRD-Crystal Pretraining (CPCP) module. The CPCP module takes PXRD patterns and crystal structures as inputs and produces similarity scores between them. A transformer-based PXRD feature extractor processes the PXRD pattern data, while an equivariant Graph Neural Network (GNN) extracts features from the crystal structure. The similarity score is computed as the dot product of these two feature sets. \textbf{b}, Framework of the Conditional Crystal Structure Generation (CCSG) module. The CCSG module utilizes the PXRD pattern as a condition to generate crystal structures. The PXRD feature extractor is initialized from the CPCP module pretraining and kept frozen. Subsequently, the composition of the crystal are used to initialize the atom positions and lattice matrix. The denoising network, referred to as the crystal structure network, takes the previous crystal structure, along with the PXRD feature obtaind through PXRD feature extractor and time step, as inputs to update the crystal structure. This process is iteratively repeated in a reverse manner. \textbf{c}, PXRD data generation workflow. PXRD data can be acquired in two ways: by simulating PXRD data from a given crystal structure using GSAS~\cite{gsas} software, or by conducting actual PXRD experiments with an XRD instrument. \textbf{d}, Workflow of the PXRD feature extractor. The PXRD data is first tokenized into peak tokens based on peak intensity and the corresponding diffraction angle, after which PXRD features are derived using BERT. \textbf{e}, Framework of the crystal structure network. In the CPCP model, only the solid line components of the process are executed, whereas in the CCSG model, both solid and dashed line components are executed.}
		\label{fig:architecture}
\end{figure}

XtalNet is designed to predict crystal structures from PXRD pattern, which we approach as a conditional generation task. The goal is to generate the corresponding crystal structure based on the given PXRD pattern. To achieve this, we have developed two key modules, as depicted in \textbf{Figure~\ref{fig:architecture}}a and b: the Contrastive PXRD-Crystal Pretraining (CPCP) module and the Conditional Crystal Structure Generation (CCSG) module. The CPCP module primarily aligns the crystal space with the PXRD space, and the pre-trained PXRD feature extractor is subsequently used to initialize the CCSG module. The CCSG module is specifically designed to reconstruct the crystal structure from a given PXRD pattern using the pre-trained PXRD feature extractor derived from the CPCP module by using equivariant diffusion model.

As shown in Figure~\ref{fig:architecture}a, the CPCP module is inspired by the design of the CLIP model~\cite{CLIP}. It employs a transformer-based~\cite{vaswani2017attention} PXRD feature extractor to encode PXRD patterns into feature representations. A crystal structure feature network, based on Equivariant Graph Neural Networks~\cite{diffcsp}, is then utilized to extract crystal structure features. The similarity between PXRD features and crystal structure features is calculated using cosine similarity. Matching pairs of PXRD patterns and corresponding crystal structures are treated as positive pairs, while non-matching pairs are considered negative. Contrastive learning is performed using the InfoNCE loss function. The framework of CCSG module is shown in Figure~\ref{fig:architecture}b. The CCSG module operates within a diffusion-based framework, where the PXRD feature serves as a key condition for generation. The reverse processes is used for inference, which aims to update the noised crystal structure. Hence the noised crystal structure, PXRD feature and time embedding are all passed through the crystal structure network to denoise the fractional coordinates and lattice. The diffusion process enables the CCSG training. Noise is added to the crystal structure via the diffusion process, and the goal of the crystal structure network is to predict this noise. It is worth mentioning that the PXRD feature extractor is initialized by the pretraining of CPCP module, which is vital for generation performance. While multiple crystal structure candidates can be sampled from CCSG module, the CPCP module can be used to rank them by the similarity between PXRD pattern and candidates. This process enables us to screen multiple candidates effectively.

The architecture of the PXRD feature extractor is illustrated in Figure~\ref{fig:architecture}d. Initially, the PXRD pattern is tokenized based on its peaks, and a [CLS] token, representing the global PXRD feature, is added at the header of perak tokens. A BERT model~\cite{devlin2018bert} is used to obtain the PXRD features, with the global feature from the header being utilized in both the CPCP and CCSG modules. The crystal structure network is employed in both modules, though with some differences, as shown in Figure~\ref{fig:architecture}e. In the CPCP module, only the composition is used as the node feature. However, in the CCSG module, the PXRD feature and time embedding are also used as node features, concatenated with the composition embedding. The output of the crystal structure network in the CPCP module is the final graph feature, used for contrastive learning, while in the CCSG module, the output includes both the coordinate score and lattice noise. The coordinate score is obtained from the final node feature after MPNN processing, and the lattice noise is derived from the final graph feature and the original lattice. The crystal structure networks in the two modules do not share parameters and are each trained independently from scratch. More details are discussed in Method~\ref{method}.

To accomplishing XtalNet training, we must obtain pairs of crystal structure and PXRD. The entire data workflow is illustrated in Figure~\ref{fig:architecture}c. While crystal structure is prevalent in various databases, PXRD patterns are not ordinarily included in standard databases. Since high-quality experimental data is extremely rare and costly, we simulate PXRD patterns from known crystal structures using GSAS~\cite{gsas} software as our training data, which can generate a large amount of data at limited cost. At the same time, we also use X-ray diffraction instruments to collect a limited number of experimental PXRD patterns to test XtalNet performance in practice as a case study.

\subsection{Dataset Preparing and Evaluation Metrics}
We curate two MOFs datasets, namely hMOF-100 and hMOF-400, based on hypothetical MOFs (hMOFs) database~\cite{hmof}, according to the atom number in the unit cell. Following the Uni-MOF~\cite{wang2023metal} splitting, we filter each split to retain only materials with 100 or fewer atoms in the unit cell, constructing train (73,332), validation (9,117), and test (9,081) sets for the hMOF-100 dataset. Similarly, we filter materials with 400 or fewer atoms in the unit cell to construct train (109,836), validation (13,730), and test (13,729) sets for the hMOF-400 dataset. Given that the original hMOFs database lacks PXRD patterns, we calculate the simulated PXRD patterns for each crystal in the dataset using GSAS~\cite{gsas}. Each PXRD pattern is simulated with a 2$\theta$ diffraction angle ranging from 3$^\circ$ to 30$^\circ$ and a step size of 0.02$^\circ$. For the structures within our dataset, diffraction angles above 30$^\circ$ produce a very small response, so the maximum diffraction angle for our simulated data is 30$^\circ$. Considering that only the relative intensities of PXRD patterns convey practical significance, we normalize the intensity of each pattern by dividing it by its maximum value. For experimental PXRD data, we use HighScore software~\cite{degen2014highscore} to determine background and do some preprocessing. Mercury~\cite{macrae2020mercury} and Crystal Toolkit~\cite{horton2023crystal} software are both used for visualize the crystal structure.
 
In order to evaluate the CPCP module, the database retrieval task is designed, which aims to identify matching crystal structures from a given set of candidate structures based on known PXRD pattern. As the purpose of CPCP module is to obtain matching relationship between PXRD space and crystal structure space, the database retrieval task is suitable for validating it.  Hence, We employ the top-$k$ hit rate as the evaluation metric, which measures the frequency at which the desired crystal structures are discovered. 
For the evaluation of CCSG module, following previous works~\cite{cdvae, diffcsp}, we assess the crystal structure using Match Rate and RMSE. Specifically, the Match Rate denotes the proportion of matched generated structures relative to all ground truth. The StructureMatcher class in pymatgen~\cite{pymatgen} is utilized. The RMSE is computed between the matched structure and ground truth, normalized by $\sqrt[3]{V/N}$, where $V$ and $N$ represent the lattice volume and atom numbers in the unit cell, respectively. The RMSE can also be obtained through the StructureMatcher class in pymatgen.

\subsection{CPCP Module Aligns PXRD Space with Crystal Structure Space}

\begin{figure}[htp]
    \centering
    \includegraphics[width=0.95\textwidth]{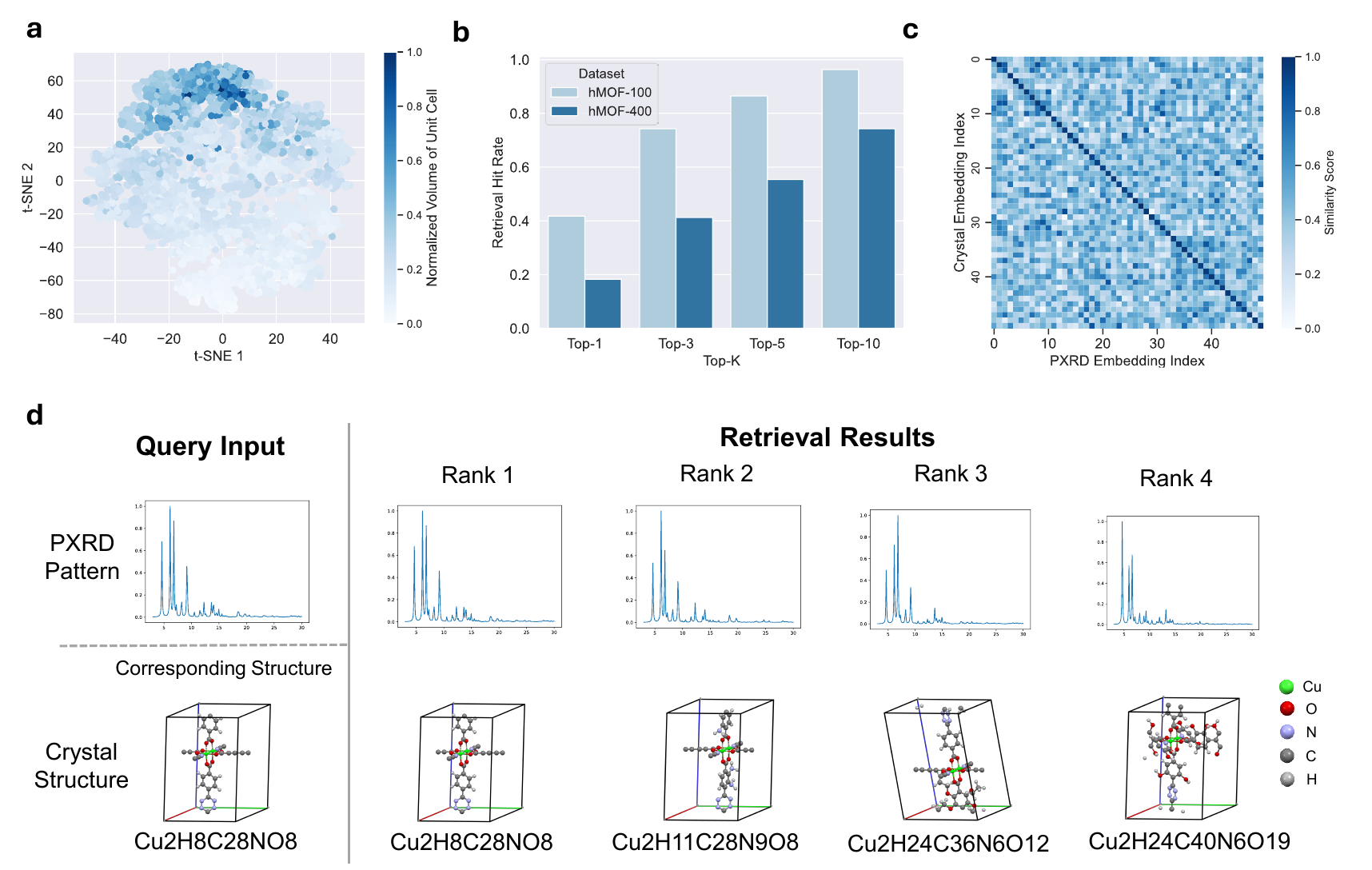}
    \caption{ \textbf{CPCP Module Performance. a}, t-SNE reduction of the hMOF-100 dataset's PXRD feature embeddings, with the unit cell volume represented by color intensity. The clustering of unit cell volume in PXRD feature embedding  indicates the effectiveness of the CPCP module in aligning PXRD and crystal structure spaces. \textbf{b}, the top-10 hit rate for the database retrieval task, highlighting the module's efficacy in identifying corresponding crystal structures based on PXRD patterns. \textbf{c}, a heatmap of similarity scores for 50 randomly selected crystal structures from hMOF-100 dataset and their corresponding PXRD patterns. \textbf{d}, a retrieval result for a given PXRD pattern, showcasing the top four retrieved crystal structures and their corresponding PXRD patterns, illustrating the high degree of similarity in metal-connecting structures and PXRD spectra.d}
    \label{fig:CPCP}
\end{figure}

To qualitatively assess the efficacy of the embeddings obtained from the CPCP module, we employ t-SNE to reduce the PXRD feature embeddings of hMOF-100 dataset into two dimensions, which is shown in \textbf{Figure~\ref{fig:CPCP}}a. The normalized volume of the unit cell corresponding to the PXRD crystal structure is represented in blue, with colors closer to blue indicating larger volumes and colors closer to white indicating smaller volumes. The clustering of colors demonstrates that similar PXRD embeddings share similar volumes. As the unit cell volume is a representative property of crystal structures, the PXRD embeddings effectively capture underlying structural characteristics, thereby aligning the PXRD space with the crystal structure space to a certain degree.

To quantitatively evaluate the alignment between PXRD and crystal structure spaces, we devise a database retrieval task to search for crystal structures based on a known PXRD pattern. Initially, both a PXRD pattern and a database containing numerous crystal structures are provided. The goal is to use the embedding derived from the PXRD pattern to identify and retrieve the most closely corresponding crystal structure.

More specifically, structure embeddings from the test sets are compiled by the crystal structure network to create a retrieval database. Searches are conducted using PXRD embeddings produced by the PXRD feature extractor as query input, employing cosine similarity as the search metric. All experiments are conduct on the whole test set of hMOF-100 and hMOF-400 dataset. The search results are summarized in Figure~\ref{fig:CPCP}b. The top-10 hit rate reaches 97.2\% in the hMOF-100 dataset and 74.1\% in the hMOF-400 dataset. Here is a performance decline from hMOF-400 to hMOF-100, which can be attributed to two main reasons. First, the model faces a more complex system, resulting in a larger search space and an increased difficulty in making accurate predictions, leading to reasonable poorer performance. Second, there is relatively less data available for larger systems compared to smaller ones, and the model has not been sufficiently trained on the large system data, which also contributes to worse results.

To gain a more intuitive understanding of the retrieval effectiveness, we randomly select 50 crystal structures and their corresponding PXRD patterns, using their similarity scores to construct a heatmap as displayed in Figure~\ref{fig:CPCP}c. In the heatmap, colors closer to blue indicate higher similarity, while colors closer to white signify lower similarity. The heatmap's diagonal represents the correctly matched pairs. A distinct blue distribution along the diagonal demonstrates the model's ability to accurately identify crystal structures corresponding to the PXRD patterns.

Additionally, we visualize a retrieval result for a given PXRD pattern. The four highest-scoring crystal structures retrieved by the CPCP module, their corresponding PXRD patterns, and their ranks are presented in Figure~\ref{fig:CPCP}d. It is evident that the non-ground-truth structures with top ranks exhibit similar metal-connecting structures, with the connected ligands also displaying a high degree of similarity. From the perspective of PXRD spectra, the peak positions of the high-intensity peaks in the PXRD spectra associated with similar structures also exhibit high similarity. Consequently, our retrieval model effectively extracts high-level information corresponding to the structures from the PXRD spectra, laying the foundation for the subsequent generation model.

\subsection{XtalNet Can Predict Crystal Structure Conditioned on PXRD}

In pursuit of generating superior corresponding crystal structures from PXRD, it is advantageous to produce a multitude of candidate structures, subsequently assigning them reasonable ranks. As the diffusion generation process includes randomness, more candidates indicate more chance to include precise crystal structure. In this study, we generate 20 crystal structure candidates for the subsequent ranking process. Utilizing the previously mentioned CPCP module, we calculate the similarity score between the target PXRD pattern and the generated crystal structures, thereby facilitating their ranking. The match rate is determined among the top-k ranking candidates, while the RMSE corresponds to the best candidate within the top-k ranking group. 

\begin{figure}[htbp]
    \centering
    \includegraphics[width=0.95\textwidth]{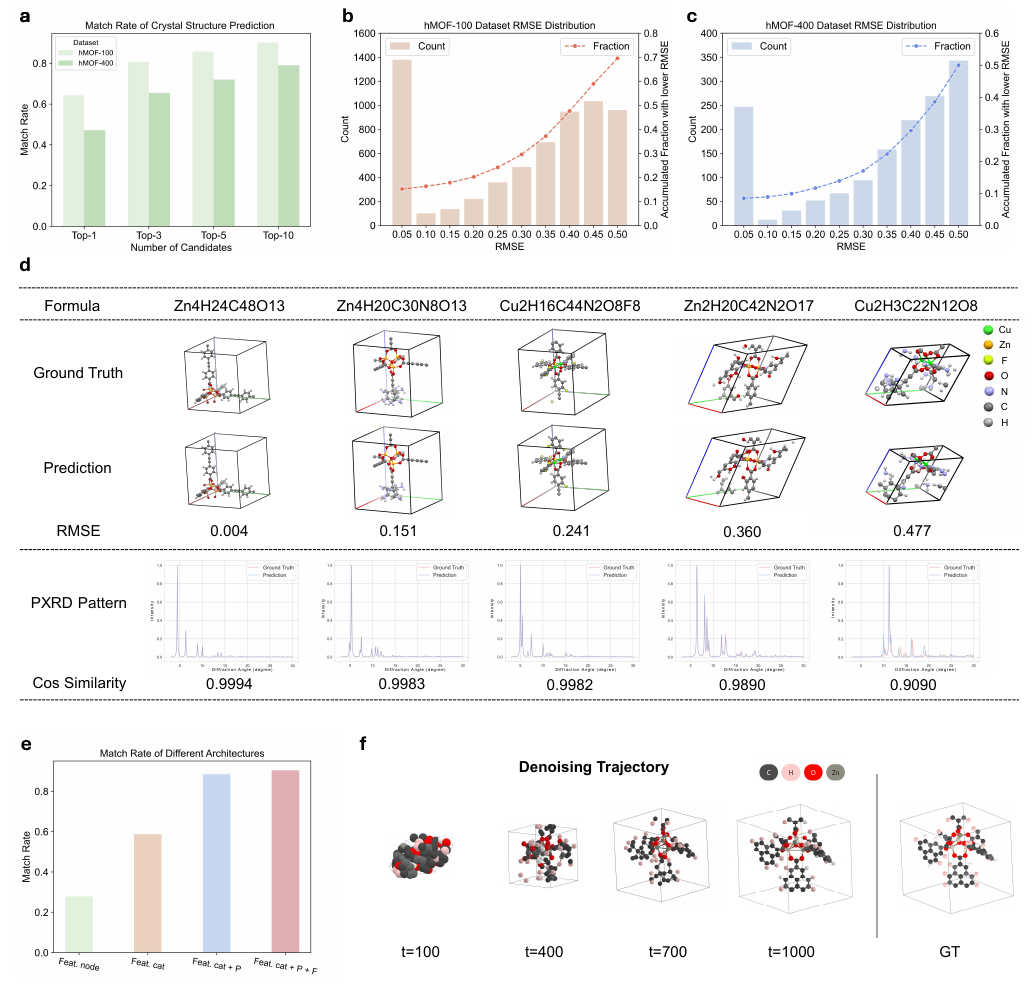}
    \caption{\textbf{Performance of XtalNet in Crystal Structure Prediction. a}, the match rates for hMOF-100 and hMOF-400 datasets with different number of top rank generated crystal structure candidates. \textbf{b,c},the RMSE statistics for hMOF-100 and hMOF-400 dataset, indicating that XtalNet can generate highly accurate crystal structures from PXRD data for a significant proportion of cases. \textbf{d}, visual comparison of generated crystal structures and their simulated PXRD patterns against ground truth, highlighting the model's performance in generating metal-connecting parts of MOFs and maintaining high similarity in PXRD patterns. \textbf{e}, performance of different architectures, demonstraing the reasonableness of XtalNet. Feat. node denotes PXRD feature is added as a new node, Feat. cat denotes PXRD feature is concated with original node features,  P denotes PXRD feature extractor is pretrained by CPCP module and F denotes PXRD feature extractor is frozen during CCSG training. \textbf{f}, diffusion trajectory of generating a crystal structure, showing the interpretability of the denoising process.}
    \label{fig:gen_per}
\end{figure}

The assessment of the hMOF-100 dataset is performed on the entire test set, while the evaluation of the hMOF-400 dataset is conducted on 3220 samples, randomly chosen from the test set, owing to the computational resources constraint. As shown in \textbf{Figure~\ref{fig:gen_per}}a, the match rates of hMOF-100 and hMOF-400 both increase as the number of candidates increases, which implies the multiple generations are useful for obtaining precise structure. The top-10 match rate of hMOF-100 dataset and hMOF-400 dataset also achieves 90.2\% and 79\% respectively. 

 The RMSE statistics of two dataset is shown in Figure~\ref{fig:gen_per}b and c. Here we use RMSE of best generation results of each test sample as the statics data.  The cases of RMSE lower than 0.05 occupy quit a bit proportion comparing with other RMSE range cases, which means XtalNet can generate very precise crystal structure from PXRD for these cases. At the same time, the proportion of samples' RMSE lower than 0.5 accounts for more than half of the total, while the RMSE of 0.5 represents the results achieves acceptable level.

 To provide a more intuitive understanding of the model's generative capabilities, we select five generated samples for visualization. We choose samples with varying RMSE to better illustrate the impact of different RMSE values on the quality of the generated results. The crystal structures and simulated PXRD pattern of ground truth and generation result for these samples are displayed in Figure~\ref{fig:gen_per}d. As the RMSE increases, the generated structure and corresponding PXRD pattern more and more deviate from ground truth. It is apparent that the model performs well in generating the metal-connecting parts of the MOFs. The generated ligand structures are generally consistent with the ground truth structures, although the finer details is not perfect. On the other hand, the PXRD patterns between generation and GT also keep high similarity. Although the peak intensity is not exactly same, the positions of the peak are almost same. 

\subsubsection{Effective Integration of PXRD in Crystal Generation}\label{diff_arch}

Incorporating PXRD features into the crystal structure generation process is non-trivial, as there are multiple potential designs that can result in varying task performance levels. As illustrated in Figure~\ref{fig:gen_per}e, we compare the match rate of different designs in hMOF-100 dataset, as the match rate is a representative metric of the generation task. Here we randomly select 1000 samples from test set to evaluate the performance. The Feat. node denotes that PXRD features are added as a new node in the crystal structure network, while Feat. cat indicates that PXRD features are concatenated with all crystal nodes. The P symbolizes that the PXRD feature extractor of the CCSG module is initialized from the CPCP module pre-training, and F signifies that the PXRD feature extractor of the CCSG module is frozen. Our strategy achieves the best results in comparison.

For the integration of PXRD features, adopting the concatenation method is more intuitive. The reason is that if PXRD features are added as a new node in the crystal structure feature extraction network, the PXRD node would possess a vastly different feature space compared to the other atomic nodes. This could result in difficulties in network learning and lead to poor performance. However, the PXRD feature extractor pretrained by the CPCP module is already well-aligned with the PXRD space and exhibits robust feature extraction capabilities, thus enhancing the performance of our model when used as initialization. Furthermore, the crystal structure generation task often requires more explicit PXRD features. Freezing the PXRD feature extractor ensures that its feature space remains unscathed by gradients during the training process.

\subsubsection{Hierarchical Optimization in the CCSG Module Generation Process}

The generation process in our CCSG module is derived from the denoising process, which can be interpreted as a continuous update of atomic positions and lattice matrix. To demonstrate the interpretability of the denoising process, we visualize the crystal structure of the same sample at different diffusion time steps in Figure~\ref{fig:gen_per}f.

As depicted in the figure, the unit cell progressively expands from a small size to ultimately resemble the ground truth (GT) unit cell shape. Concurrently, the atoms initially form coarse-grained clusters (t=400), followed by the optimization of fine-grained atomic positions (t=700), and eventually yield the generated output (t=1000). This demonstrates the hierarchical optimization strategy employed in the generation process.

\subsubsection{Evaluation of XtalNet in Diverse System Sizes and Elemental Compositions}

\begin{figure}[tbp]
    \centering
    \includegraphics[width=0.95\textwidth]{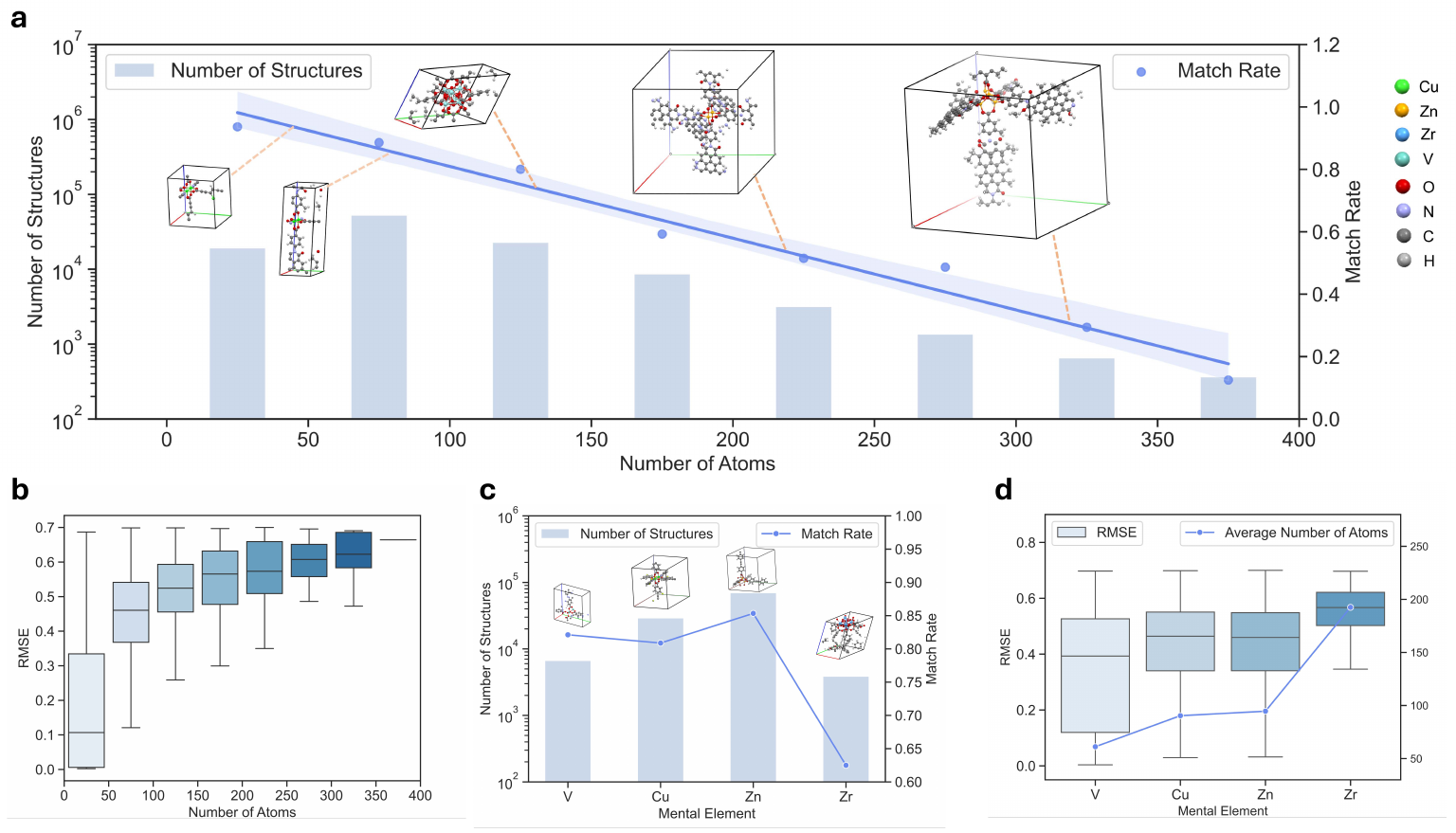}
    \caption{\textbf{Evaluation of XtalNet in Diverse System Sizes and Elemental Compositions. a}, the match rate and structure number corresponding to different system sizes in the training set, demonstrating XtalNet's applicability to systems with varying atom numbers in the unit cell. \textbf{b},  the RMSE of the best generation results for different system sizes in the hMOF400 dataset, revealing the impact of system complexity on prediction accuracy. \textbf{c and d}, the match rates and RMSE for structures containing distinct metal elements, highlighting the influence of sample number and system complexity on model performance. }
    \label{fig:len_type}
\end{figure}

XtalNet can be applied to systems with varying numbers of atoms in the unit cell, effectively accommodating diverse system sizes. The match rate and structure number corresponding to different system sizes in the training set are depicted in \textbf{Figure~\ref{fig:len_type}}a, alongside visualizations of various system sizes. The RMSE of best generation result for different system sizes is illustrated in Figure~\ref{fig:len_type}b. Here All results are derived in hMOF-400 dataset. As the system size increases, the match rate decreases, and the RMSE also escalates. This trend can be attributed to two factors: the increasing complexity of structure generation as the system size expands, and the reduced number of larger system structures, potentially leading to imbalanced and insufficient training.

XtalNet is also applicable to systems containing different metal elements. The hMOF-400 dataset primarily comprises four metal element types: V, Cu, Zn and Zr. The match rates and RMSE for structures containing distinct metal elements are displayed in Figure~\ref{fig:len_type}c and d. It can be observed that the RMSE increases with the average number of atoms in the unit cell, indicating that more complex crystal structures yield poorer prediction results, which aligns with general expectations. Notably, the match rate for crystals containing Zn is significantly higher than those with Cu, which may be due to the larger number of samples containing Zn. An increased sample size can result in more comprehensive training and optimization. Consequently, the performance is influenced by both the number of samples and the complexity of the system.

\subsection{Application of XtalNet to Real Experimental PXRD Patterns}

\begin{figure}[tbp]
    \centering
    \includegraphics[width=0.98\textwidth]{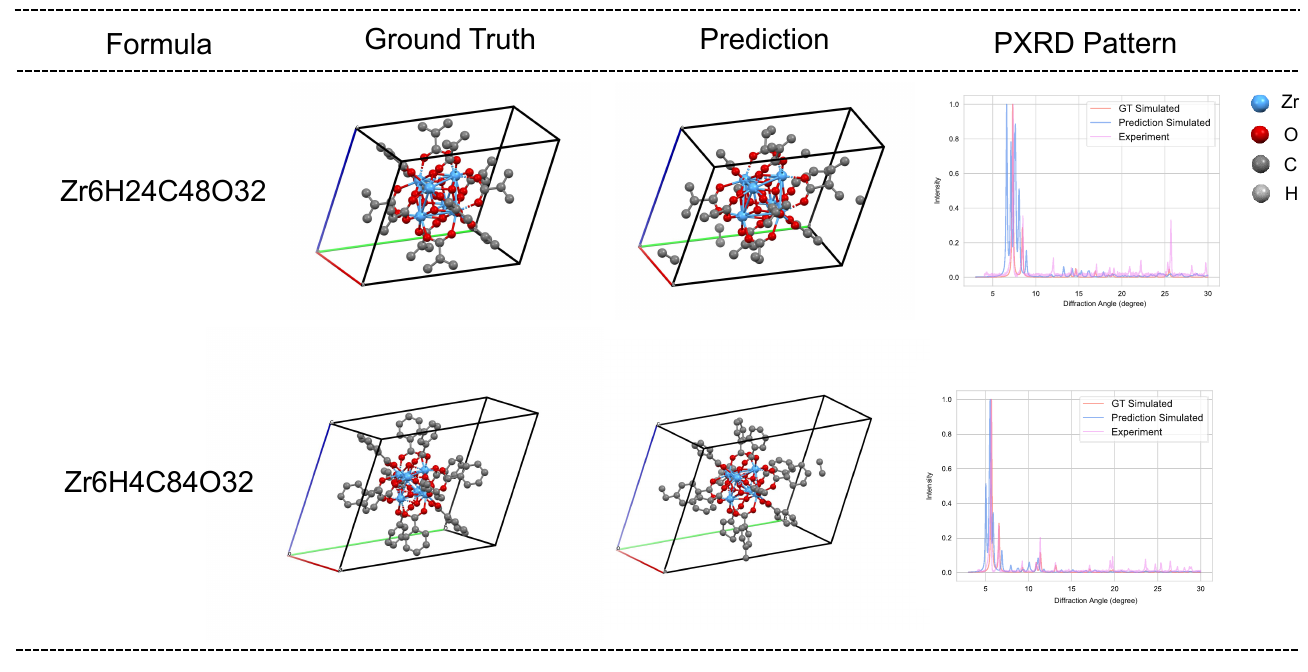}
    \caption{\textbf{XtalNet Predictions of Real Experimental PXRD Patterns.} Two cases of XtalNet's predictions for real experimental PXRD data are drawn, showcasing both the ground truth (GT) crystal structures and the predicted crystal structures. The GT simulated (red), predicted simulated (blue), and experimental (purple) PXRD patterns are also presented for comparison.}
    \label{fig:real}
\end{figure}

The paramount objective of 
XtalNet is its application to experimental PXRD data. Owing to the considerable discrepancy between simulated and experimental PXRD data, coupled with the presence of multiple materials in experimental samples, predicting crystal structures from experimental PXRD patterns poses a formidable challenge. Nevertheless, 
XtalNet exhibits fair performance, thereby highlighting its robustness. As illustrated in \textbf{Figure~\ref{fig:real}}, two results of real experimental PXRD are depicted. A high degree of similarity is observed between the predicted and ground truth (GT) structures, particularly in the metallic components. The experimental PXRD pattern exhibits minor noise and slight differences in the high diffraction angle as compared to the GT simulated PXRD. For the predicted PXRD pattern, the majority of peak positions exhibit similarity with the experimental PXRD, which is crucial for structure determination. Several conspicuous noise peaks are also evident in the predicted simulated PXRD, potentially attributable to minor errors in ligand positioning that may alter the symmetry of the overall structure. Overall, XtalNet is applicable in predicting crystal structure from real experimental PXRD pattern.

\section{Discussion}
The development and implementation of XtalNet represent a significant leap forward in the field of both crystal structure prediction and automated PXRD data analysis. Our model, which employs an end-to-end deep learning framework, has demonstrated the capability to accurately predict crystal structures that match given PXRD patterns without reliance on external databases. This is a substantial departure from traditional methods that often require extensive manual intervention and database matching, which can be time-consuming and yield suboptimal results due to the incompleteness of database coverage. 

XtalNet is a method suitable for predicting any crystal structure system, and it does not impose any requirements on the system itself. Therefore, as long as there is a corresponding reasonable dataset, it can be used for both organic and inorganic systems. Here, we chose the more challenging organic system, MOF, to validate our method. The success of XtalNet in generating crystal structures underscores the model's ability to handle complex systems such as metal-organic frameworks (MOFs). The contrastive learning approach and diffusion-based conditional generation used in XtalNet have proven effective in establishing a one-to-one mapping between the PXRD space and crystal structure space, thereby reducing ambiguity in the prediction of multiple stable crystal structures for a given chemical composition.

Due to the challenges associated with collecting PXRD experimental data, there is indeed a scarcity of such data. To address this limitation, we utilized GSAS software to simulate the PXRD data, generating a relatively large dataset for training purposes. This approach effectively mitigated the impact of data scarcity and resulted in promising outcomes. The high top-10 hit ratio and match rate indicate that XtalNet can effectively retrieve and generate crystal structures that closely match PXRD patterns, which is crucial for applications in material science as precise structural information is essential for understanding material properties and guiding the design of new materials with tailored characteristics.

However, despite these advancements, there are some limitations to the current study that warrant further exploration. XtalNet has demonstrated impressive results in generating crystal structures from simulated PXRD data, but the application of this model to real experimental PXRD data presents a unique set of challenges. The complexity and variability inherent in experimental data, such as noise, peak broadening, effect of solvent and preferred orientation, can significantly impact the accuracy and reliability of the crystal structure predictions made by XtalNet. Consequently, the application of deep learning techniques under complex experimental conditions remains a challenge. In the future, more data augmentation and  simulation data enhancement can be taken to improve the performance.

\section{Methods}\label{method}
The unit cell constitutes the fundamental repeating unit that serves as the foundational building block for characterizing 3D crystal structures. A unit cell can be defined as $\gM=(\mA, \mF, \mL)$, where $\mA \in \sR^N$ represents the atom types, $\mF \in [0,1)^{3 \times N}$ denotes the fractional coordinates of the atoms within the unit cell, and $\mL \in \sR^{3 \times 3}$ is the lattice matrix representing the periodicity of the crystal, defined by three basis vectors derived using the Niggli algorithm~\cite{santoro1970determination}, and $N$ signifies the number of atoms in each unit cell.For practical crystal identification problems based on PXRD, we can assume that the PXRD pattern $\mC_{XRD}$, the atom types $\mA$, and the number of atoms $N$ are provided. In our setup, conditional crystal structure prediction (CSP) aims to model the conditional distribution $p(\mL, \mF \mid \mA, \mC_{XRD})$, learning the relationship between the crystal lattice, fractional coordinates, atom types, and PXRD patterns to predict the most probable crystal structure.

In our approach, we utilize paired PXRD and crystal structure data during the training phase. This method incorporates two specialized neural networks: a PXRD Feature Extractor, which derives essential features from PXRD patterns, and a Crystal Structure Network, which extracts structural features from given crystal structures and predicts their subsequent states. To effectively train these networks, we adopt a dual-task framework. The first task involves aligning the PXRD features produced by the PXRD Feature Extractor with the crystal structure features obtained from the Crystal Structure Network. This task employs a contrastive learning strategy, akin to the approach utilized by CLIP~\cite{CLIP}, aimed at aligning data across different modalities. The second task focuses on generating crystal structures using the Crystal Structure Network, based on features predicted by the PXRD Feature Extractor.

\subsection{Model Achitecture}

\paragraph{PXRD Feature Extractor}

To identify patterns in Powder X-ray Diffraction (PXRD) data, we developed a neural network called the PXRD Feature Extractor ($f_{PXRD}$), which is based on the BERT architecture~\cite{devlin2018bert}. PXRD data are represented as curves with diffraction angles on the x-axis and intensities on the y-axis. Before inputting the PXRD data into the model, we preprocess the data by normalizing the intensities and smoothing the curves to reduce noise. This step ensures that the input data is standardized, facilitating better learning and reducing the impact of experimental variability.

Given that peaks in PXRD data are critically important, our approach focuses specifically on these peaks, characterized by their specific 2$\theta$ angles $\mA_{xrd}\in\sR^{L}$ and intensity magnitudes $\mI_{xrd}\in\sR^{L}$, where $L$ is the number of peaks. Focusing on peak data provides a more concise representation of the PXRD pattern, which is essential for computational efficiency. Peaks represent the most informative aspects of PXRD data, corresponding to the Bragg reflections that directly relate to the crystal structure. By concentrating on these key features, we effectively reduce the length of the data sequences, optimize model performance, and maintain the most relevant information for crystal structure prediction. Moreover, using peak data instead of the full continuous spectrum allows us to focus on the most distinctive features of the diffraction pattern, which are crucial for differentiating between similar crystal structures. This approach aligns with traditional crystallographic analysis, where peak positions and intensities are often sufficient to determine the phase and structure of a material. While the full continuous spectrum contains more detailed information, the peak-focused approach strikes a balance between detail and computational efficiency, enabling the model to generalize better across different datasets.

To preprocess the PXRD data for model input, $\mA_{xrd}$ is treated as discrete positions indicative of peak locations, which are converted into positional embeddings using an embedding layer, similar to the position encodings in traditional Transformer models. Since $\mI_{xrd}$ represents continuous intensity values, we use a multilayer perceptron (MLP) network to transform these values into input embeddings. Additionally, to enhance the representation capability of the model, a unique trainable embedding is added at the beginning of the tokenized sequence, similar to the "[CLS]" token in BERT models. This embedding serves as a summary representation, or "header," for the PXRD data. The PXRD header feature, $\mP = f_{PXRD}(\mA_{xrd}, \mI_{xrd})$, is subsequently used as the comprehensive representation of the entire PXRD data for contrastive pretraining and as a conditioning input for the diffusion models. The PXRD feature extractor is pre-trained in the Contrastive PXRD-Crystal Pretraining (CPCP) module to align PXRD features $\mP$ with crystal structure feature $\mC$. During the training of the Conditional Crystal Structure Generation (CCSG) module, the PXRD feature extractor is initialized with pre-trained parameters and remains frozen to maintain the learned feature representations.

\paragraph{Crystal Structure Network}

In our study, we utilize a Modified Equivariant Graph Neural Network (EGNN), referred to as CSPNet~\cite{diffcsp}, derived from the DiffCSP framework, to serve as the Crystal Structure Network ($f_{CSP}$). This network is specifically tailored for handling crystallographic data. To align with our specific research goals, several modifications were made to the original CSPNet. For the diffusion process, CSPNet constructs node features by initially combining atom attributes (such as atom types) with time embedding. These combined features are then transformed to produce final node representations through Message Passing Neural Networks (MPNN). Additionally, we introduce a novel conditioning signal, incorporating temporal embeddings and condition-specific PXRD features $\mP$, to better guide the denoising process. This is achieved by integrating node features with both time embedding and condition-specific PXRD features $\mP$ before passing them through a linear transformation layer:

$$
\mN^0 =  \left\{
\begin{aligned}
\text{Linear}(\text{cat}(\mP, \text{Embedding}(\mA), t)), \text{in CCSG module} \\
\text{Embedding}(\mA), \text{in CPCP module}
\end{aligned}
\right.
$$

where $\mN^0$ denotes the initial node feature, $\mP$ denotes the PXRD header feature, $t$ represents the time embedding. The directional information between atoms is processed by Fourier transformation as the edge features. Subsequently, the node features, edge features and lattice parameters are all passed through MPNN to obtain final node features and global graph feature. The single-layer structure of MPNN is as follows:
\begin{align*}
    m_{i,j}^k &=  \text{MLP}(\mN_{i}^{k}, \mN_{j}^{k}, \mL^\top\mL, \text{FT}(\mF_i-\mF_j)) \\
    m_i^k &= \sum_j m_{i,j}^k \\
    \mN_i^{k+1} &= \mN_i^{k} + \text{MLP}(\mN_i^{k}, m_i^k)
\end{align*}

where $\mN_i^k$ denotes the $i$th node feature of $\mN^k$ in $k$th layer, $\mF_i$ denotes the $i$th node(atom) coordinate of $\mF$ and $\text{FT}(\cdot)$ denotes Fourier transformation. After $s$ layers processing, the global graph feature and the denoising output can be derived as:
\begin{align*}
    \mC & = \text{Linear}(\frac{1}{N}\sum_i^N m_i^s) \\
    \hat{\varepsilon}_\mF^i &= \text{MLP}(\mN_i^s) \\
    \hat{\varepsilon}_\mL &= \mL\cdot\text{MLP}(\frac{1}{N}\sum_i^N m_i^s)
\end{align*}

where $\mC$ denotes the global crystal feature, $\hat{\varepsilon}_\mF^i$ denotes the $i$th column of the denoising socre $\hat{\varepsilon}_\mF$ for fractional coordinates, and $\hat{\varepsilon}_\mL$ denotes the denoising term for the lattice. Note that in the CPCP module, $\mP$, $t$, $\hat{\varepsilon}_\mF$, and $\hat{\varepsilon}_\mL$ are not used.

In CPCP module, the crystal feature $\mC$ is used as the output of crystal structure network for contrastive learning, which captures the aggregated structural characteristics of the crystal. In CCSG module, $\hat{\varepsilon}_\mF$, and $\hat{\varepsilon}_\mL$ are the output of crystal struture network. Additionally, $\mP$ from the the pre-trained PXRD feature extractor provides conditional PXRD information that guides the denoising direction of the crystal structure. The PXRD header feature serves as a guide throughout reverse process, ensuring the synthesized crystal structure is not only in agreement with the PXRD data but also conforms to the material system's intrinsic physical and chemical properties. This strategic enhancement integrates critical supervisory data into the model's architecture.

\subsection{Contrastive PXRD-Crystal Pretraining (CPCP) Module}

The Contrastive PXRD-Crystal Pretraining (CPCP) module aims to align the embedding spaces of PXRD patterns and crystal structures to facilitate a better understanding of their underlying relationships. In this module, the PXRD feature extractor generates a PXRD header feature vector $\mP$ from the PXRD data, and the Crystal Structure Network computes a crystal structure feature vector $\mC$ from the crystal graph representation.

To train these embeddings, we use a contrastive learning approach that involves constructing both positive and negative pairs. Positive pairs are formed by pairing PXRD features $\mP_i$ and crystal structure features $\mC_i$ that correspond to the same material. Negative pairs, on the other hand, are constructed by pairing a PXRD feature $\mP_i$ with a crystal structure feature $\mC_j$ from a different material ($i \neq j$). This setup forces the model to learn robust embeddings by minimizing the distance between positive pairs (similar materials) while maximizing the distance between negative pairs (dissimilar materials).

Constructing negative pairs is crucial as it encourages the model to differentiate between features of different materials, enhancing the model's discriminative power. The contrastive learning task helps bridge the gap between PXRD data and crystal structures by aligning their representations in a shared embedding space. This alignment allows the model to learn more meaningful correlations between diffraction patterns and their corresponding crystal structures, improving its ability to predict crystal structures from PXRD data.

The loss function used for this task is the InfoNCE loss, which optimizes the embeddings by encouraging high similarity for positive pairs and low similarity for negative pairs. The \textbf{Contrastive Learning Loss} is defined as:

\begin{align}
    \gL_{c} = -\sum_{i=1}^N \log \frac{e^{\text{sim}(\mP_i, \mC_i) / \tau}}{\sum_{j=1}^N e^{\text{sim}(\mP_i, \mC_j) / \tau}} -\sum_{i=1}^N \log \frac{e^{\text{sim}(\mC_i, \mP_i) / \tau}}{\sum_{j=1}^N e^{\text{sim}(\mC_i, \mP_j) / \tau}}
\end{align}

\begin{equation}
\text{sim}(\mP_i, \mC_j) = \frac{\mP_i \cdot \mC_j}{\Vert \mP_i\Vert \Vert \mC_j\Vert}
\end{equation}

where $\text{sim}(\mP_i, \mC_j)$ is the cosine similarity between PXRD and crystal structure embeddings, and $\tau$ is the temperature parameter controlling the sharpness of the distribution.

By using this approach, the model learns a shared embedding space where PXRD and crystal structure features are meaningfully aligned, thus improving its ability to interpret and predict crystal structures from PXRD data.

\subsection{Conditional Crystal Structure Generation (CCSG) Module}

The Conditional Crystal Structure Generation (CCSG) module employs a diffusion-based approach to generate crystal structures that are consistent with input PXRD data. The Crystal Structure Network starts with an initial noisy crystal structure and iteratively refines it to produce a final structure that aligns with the PXRD features $\mP$. This conditioning ensures that the generated structure matches the PXRD pattern.

The forward diffusion process involves the systematic addition of Gaussian noise to the lattice parameters and fractional coordinates at each timestep, gradually increasing their randomness. Conversely, the reverse diffusion process aims to iteratively remove this noise, refining the structure towards its true form. The diffusion model is defined separately for lattice parameters $\mL$ and fractional coordinates $\mF$, with both processes described as follows:

\paragraph{Lattice Parameter Diffusion}
The diffusion process for lattice parameters $\mL$ starts with an initial Gaussian distribution:

\begin{equation} p(\mL_T) = \mathcal{N}(0, \mathbf{I}), \end{equation}

where $\mL_T$ represents the noisy lattice state at the final diffusion step $T$. The forward diffusion process progressively adds Gaussian noise to $\mL_{t-1}$ to obtain $\mL_t$:

\begin{equation} q(\mL_t | \mL_{t-1}) = \mathcal{N}(\mL_t \mid \sqrt{1 - \beta_t} \mL_{t-1}, \beta_t \mathbf{I}), \end{equation}

where $\beta_t \in (0, 1)$ is a variance control parameter that determines the noise level added at each step $t$. Cumulatively, the distribution of $\mL_t$ conditioned on the initial lattice $\mL_0$ is:

\begin{equation} q(\mL_t | \mL_{0}) = \mathcal{N}\left(\mL_t \mid \sqrt{\bar{\alpha}t}\mL_{0}, (1 - \bar{\alpha}_t)\mathbf{I}\right), \end{equation}

where $\bar{\alpha}t = \prod{s=1}^t (1-\beta_s)$, effectively capturing the aggregated effect of noise over time, often scheduled by a cosine function~\cite{nichol2021improved}.

In practice, the input typically consists of both $\mL$ and $\mF$ together, rather than just $\mL$. Therefore, the equation becomes:

\begin{equation*}
p(\mL_{t-1} | \mM_t) = \gN(\mL_{t-1} | \mu(\gM_t),\sigma^2 (\gM_t)\mI), 
\end{equation*}

where $\gM_t$ is the combination of $L_t$ and $F_t$, the mean $\mu(\mM_t)$ and variance $\sigma^2(\mM_t)$ are given by:

\begin{align} \mu(\gM_t) = \frac{1}{\sqrt{\alpha_t}}\Big(\mL_t - \frac{\beta_t}{\sqrt{1-\bar{\alpha}_t}}\hat{\varepsilon}_{\mL}\Big), \\
 \sigma^2(\gM_t) = \beta_t \frac{1-\bar{\alpha}_{t-1}}{1-\bar{\alpha}_t} \end{align}

Here, $\hat{\varepsilon}_{\mL}$ is the predicted noise term for lattice parameters, learned by the Crystal Structure Network.

\paragraph{Fractional Coordinate Diffusion}
For fractional coordinates $\mF$, the diffusion process similarly adds noise, accounting for periodic boundary conditions. The forward diffusion step for $\mF$ at time $t$ is modeled by:

\begin{equation} \mF_t = w(\mF_0 + \sigma_t \vepsilon_\mF), \end{equation}

where the truncation function $w(\cdot)$ ensures $\mF$ remains within periodic boundaries, $\vepsilon_\mF \sim \mathcal{N}(0, \mathbf{I})$ is the Gaussian noise added, and $\sigma_t$ is a noise scale parameter defined by an exponential scheduler: $\sigma_t = \sigma_1\left(\frac{\sigma_T}{\sigma_1}\right)^{\frac{t-1}{T-1}}$.

The forward distribution under this setup is given by the Wrapped Normal (WN) transition:

\begin{equation} q(\mF_t | \mF_0) \propto \sum_{\mathbf{Z} \in \mathbb{Z}^{3 \times N}} \exp\left(-\frac{|\mF_t - \mF_0 + \mathbf{Z}|_F^2}{2\sigma_t^2}\right). \end{equation}

The reverse process aims to denoise $\mF_t$ back to $\mF_{t-1}$ using an ancestral sampling approach enhanced with Langevin dynamics, where the denoising term $\hat{\varepsilon}_{\mF}$ is again predicted by the Crystal Structure Network.

By applying these forward and reverse processes to both lattice parameters and fractional coordinates, the model iteratively refines the noisy inputs to generate crystal structures that are consistent with the PXRD data.

\subsection{Diffusion-Based Generation Loss}

To optimize the CCSG module, the model minimizes the loss functions for both lattice parameters and fractional coordinates:

Lattice Loss: Minimizes the difference between predicted and true noise terms for lattice parameters:
\begin{equation} \mathcal{L}_\mL = \mathbb{E}_{\vepsilon_\mL \sim \mathcal{N}(0,\mathbf{I}), t \sim \mathcal{U}(1,T)}\left[|\varepsilon_{\mL} - \hat{\varepsilon}_{\mL}|_2^2\right]. \end{equation}

Fractional Coordinate Loss: Focuses on the denoising of fractional coordinates:
\begin{equation} \mathcal{L}_\mF = \mathbb{E}_{\mF_t \sim q(\mF_t | \mF_0), t \sim \mathcal{U}(1,T)} \left[\lambda_t|\nabla_{\mF_t}\log q(\mF_t | \mF_0)-\hat{\varepsilon}_\mF|_2^2\right], \end{equation}

where $\lambda_t = \mathbb{E}^{-1}_{\mF_t}\left[|\nabla{\mF_t}\log q(\mF_t | \mF_0)|_2^2\right]$ is approximated via Monte Carlo sampling~\cite{diffcsp}.

The Lattice Loss $\mathcal{L}_\mL$ minimizes the L2 norm of the difference between the predicted and actual noise components, effectively learning the noise pattern and improving denoising accuracy. The Fractional Coordinate Loss $\mathcal{L}_\mF$ leverages gradient information to enhance the model's sensitivity to structural variations, further refining its predictive capabilities. By optimizing these loss functions, the model learns to generate accurate crystal structures that align with PXRD data while maintaining consistency with the physical and chemical properties of the materials.

\subsection{Implementation Details}
XtalNet is trained on hMOF-100 and hMOF-400,  datasets of synthesized PXRD patterns and their corresponding crystal structures. We employ the Adam optimizer with a learning rate of 1e-4 for CPCP module training and 1e-3 for CCSG module. We use a batch size of 64 for hMOF-100 dataset in all modules, 32 for hMOF-400 dataset in CPCP module and 16 for hMOF-400 dataset in CCSG module. The networks are implemented in PyTorch and trained on NVIDIA V100 GPUs for 400 epochs with 20 epochs warmup for all datasets and modules.

\medskip
\textbf{Data Availability Statement} \par
The dataset, checkpoint and code that support the findings of this study are openly available in zenodo: \href{https://zenodo.org/records/13629658}{https://zenodo.org/records/13629658}.

\medskip
\textbf{Supporting Information} \par
Supporting Information is available from the Wiley Online Library or from the author.

\medskip
\textbf{Acknowledgements} \par
This work is supported by the National Natural Science Foundation of China (22125502), National Science and Technology Major Project (2022ZD0114902) and National Science Foundation of China (NSFC62276005).
\medskip
\textbf{Conflict of Interest} \par
The authors declare no conﬂict of interest.

\medskip

\bibliographystyle{MSP}
\bibliography{MSP-template}

\end{document}

%% file: math_command.tex
\def\vepsilon{{\bm{\epsilon}}}

\def\mA{{\bm{A}}}

\def\mC{{\bm{C}}}

\def\mF{{\bm{F}}}

\def\mI{{\bm{I}}}

\def\mL{{\bm{L}}}
\def\mM{{\bm{M}}}
\def\mN{{\bm{N}}}

\def\mP{{\bm{P}}}

\DeclareMathAlphabet{\mathsfit}{\encodingdefault}{\sfdefault}{m}{sl}
\SetMathAlphabet{\mathsfit}{bold}{\encodingdefault}{\sfdefault}{bx}{n}

\def\gL{{\mathcal{L}}}
\def\gM{{\mathcal{M}}}
\def\gN{{\mathcal{N}}}

\def\sR{{\mathbb{R}}}

